\documentclass[pdflatex,sn-mathphys-num]{sn-jnl}% Math and Physical Sciences Numbered Reference Style 
\usepackage{graphicx}%
\usepackage{multirow}%
\usepackage{amsmath,amssymb,amsfonts}%
\usepackage{amsthm}%
\usepackage{mathrsfs}%
\usepackage[title]{appendix}%
\usepackage{xcolor}%
\usepackage{textcomp}%
\usepackage{manyfoot}%
\usepackage{booktabs}%
\usepackage{algorithm}%
\usepackage{algorithmicx}%
\usepackage{algpseudocode}%
\usepackage{listings}%
\theoremstyle{thmstyleone}%
\theoremstyle{thmstyletwo}%

\theoremstyle{thmstylethree}%

\begin{document}

\title[Modelling infodemics on a global scale: A 30 countries study using epidemiological and social listening data]{Modelling infodemics on a global scale: A 30 countries study using epidemiological and social listening data}


\author[1]{\fnm{Edoardo} \sur{Loru}}\email{edoardo.loru@uniroma1.it}

\author[2]{\fnm{Marco} \sur{Delmastro}}\email{marco.delmastro@cref.it}

\author[3,4]{\fnm{Francesco} \sur{Gesualdo}}\email{francesco.gesualdo@med.unipi.it}

\author*[5]{\fnm{Matteo} \sur{Cinelli}}\email{matteo.cinelli@uniroma1.it}

\affil[1]{\orgdiv{Department of Computer, Control and Management Engineering}, \orgname{Sapienza University of Rome}, \orgaddress{\street{Via Ariosto 25}, \city{Rome}, \postcode{00185}, \country{Italy}}}

\affil[2]{\orgdiv{Department of
Environmental Sciences, Informatics and Statistics}, \orgname{Ca’ Foscari University of Venice}, \orgaddress{\street{Dorsoduro 3246}, \city{Venice}, \postcode{30123}, \country{Italy}}}

\affil[3]{\orgname{Centro Ricerche Enrico Fermi}, \orgaddress{\street{Via Panisperna 89 A}, \city{Rome}, \postcode{00184}, \country{Italy}}}

\affil[4]{\orgdiv{Department of Translational Research and New Technologies in Medicine and Surgery}, \orgname{University of Pisa}, \orgaddress{\street{Via Savi 10}, \city{Pisa}, \postcode{56126}, \country{Italy}}}

\affil*[5]{\orgdiv{Department of Computer Science}, \orgname{Sapienza University of Rome}, \orgaddress{\street{Via Regina Elena 295}, \city{Rome}, \postcode{00161}, \country{Italy}}}

\abstract{
\textbf{\textcolor{red}{Please refer to published version:}} \href{https://doi.org/10.1093/pnasnexus/pgag274}{10.1093/pnasnexus/pgag274}\\
Infodemics represent a significant threat to public health, arising from complex interactions between online and offline phenomena. The continuous feedback loops between digital information ecosystems and real-world contingencies make infodemics particularly challenging to define operationally, measure, and eventually model in quantitative terms.
This study aims to evaluate the effect of various epidemic-related variables on the dynamics of the COVID-19 infodemic, using a regression modelling framework applied to data from 30 countries across diverse income groups.
We use WHO COVID-19 surveillance data on new cases and deaths, vaccination data from the Oxford COVID-19 Government Response Tracker, infodemic data (volume of public conversations and social media content) from the WHO EARS platform, and Google Trends data to represent information demand. 
Our findings show that new deaths are the strongest predictor of document production, and that the epidemic burden in neighbouring countries exerts a greater influence on document production than domestic epidemic conditions. Building on these results, we propose a data-driven classification of country-level response that highlights country-specific discrepancies between the evolution of the infodemic and the epidemic. Further, an analysis of the temporal evolution of the relationship between the two phenomena quantifies the extent to which discussions surrounding vaccine rollouts may have shaped the development of the infodemic. Beyond underscoring the value of a holistic approach that integrates both online and offline dimensions, our results demonstrate that the evolution of infodemics and their relationship with epidemic variables can be closely monitored, even over short time windows.}

\maketitle

\section*{Introduction}\label{sec:intro}
The continuously growing amount of data and information accessible online have generated both opportunities and threats~\cite{cinelli2021echo, lorenz2023systematic, shao2018spread, vosoughi2018spread}, one of which is the risk of infodemics~\cite{zarocostas2020fight, briand2021infodemics}. An infodemic is defined as ``an excess information of varying quality, including false/misleading information or ambiguous information or both, that spreads in digital and physical environments during a health emergency"~\cite{wilhelm2023measuring}. The COVID-19 infodemic was one of the most relevant phenomena that took place in the online ecosystem during the last decade~\cite{cinelli2020covid}. According to experimental and real-world evidence, that infodemic had a number of offline consequences, such as the use of autocures~\cite{aghababaeian2020alcohol, temple2021toxic}, reduced vaccine acceptance~\cite{loomba2021measuring}, and weakened social cohesion, trust, and order~\cite{betsch2020social,jewett2021social, van2022misinformation}. Past the COVID-19 health emergency, the scientific community is still actively discussing the infodemic phenomenon, focusing on: the chance of an overflow of infodemics beyond the boundaries of public health~\cite{alipour2024cross, pagoto2023next}; the role that new technologies, such as generative AI, will play in future infodemics~\cite{de2023chatgpt}; the systematization of infodemic management frameworks~\cite{eysenbach2002infodemiology,eysenbach2020fight, ishizumi2024beyond}, often borrowed from epidemiology~\cite{scales2021covid}. While it is reasonable to assume that epidemics and infodemics are interconnected~\cite{briand2021infodemics, cinelli2025infodemic} and metaphorically similar~\cite{simon2023autopsy}, simply adapting epidemic management frameworks to infodemic management might not be sufficient.

Infodemics, as highlighted by WHO, extends beyond the spread of information to include its interplay with behavioural, societal, and health system dynamics~\cite{wilhelm2023measuring}. Public health outcomes depend on factors such as information avoidance, community-level contextualization, and systemic challenges in health communication. Although these more comprehensive dimensions are essential to fully understanding infodemics, most quantitative studies have focused on narrower aspects, often reducing the phenomenon to single measures, such as the circulation of misinformation \cite{cinelli2020covid,gallotti2020assessing} or information voids \cite{purnat2021infodemic}.
Similarly, in this study, we narrow our focus on document production as a measurable proxy for the infodemic. However, unlike prior studies, we take an additional step by developing an explanatory model to systematically analyse the factors driving infodemic dynamics related to information production and their relationship with epidemic variables. While explanatory models are commonly used in epidemiology to study disease spread~\cite{bonaccorsi2020economic, vandentorren2022effect}, their application to infodemics remains largely unexplored. To effectively tackle, manage, and prepare for future infodemics, it is paramount to address this gap by systematically analysing the factors driving their evolution.

To address this need, following the recent recommendation for the use of a holistic approach in the analysis of information ecosystems related to public health~\cite{jin2024social}, the present study proposes a methodological framework for infodemic modelling, applied to the COVID-19 epi-infodemic. 
A key difficulty is the operationalization of the definition of `infodemic' within a quantitative framework. While the World Health Organization (WHO) defines an infodemic conceptually as an `overabundance of information'--including both accurate and inaccurate claims--translating this definition into a measurable metric remains methodologically challenging. In the absence of a standardized empirical threshold for what constitutes `overabundance', we leverage data from the WHO monitoring service and use the volume of digital documents as a proxy for information abundance, modeling this phenomenon through large positive deviations in time-series data.
Specifically, we use the volume of social media content and public conversations as a proxy for the infodemic, and we then investigate its relationship with epidemic variables, namely new cases and new deaths.
We develop a global model to identify the key drivers of the COVID-19 infodemic, incorporating the effects of both epidemiological factors (new cases and deaths) and non-epidemiological factors across different areas. Our analysis includes data from 30 countries representing diverse income groups. To ensure robustness, we replicate the model using information demand, a relevant quantity in infodemic studies~\cite{mavragani2019google, purnat2021infodemic} that we measure through Google Trends data, as an alternative to information production.

\section*{Results}
\label{sec:results}

\subsection*{Epidemic and infodemic curves}
\label{sec:epi_info_curves}
\begin{figure*}[!t]
\centering
\includegraphics[width = 1\linewidth]{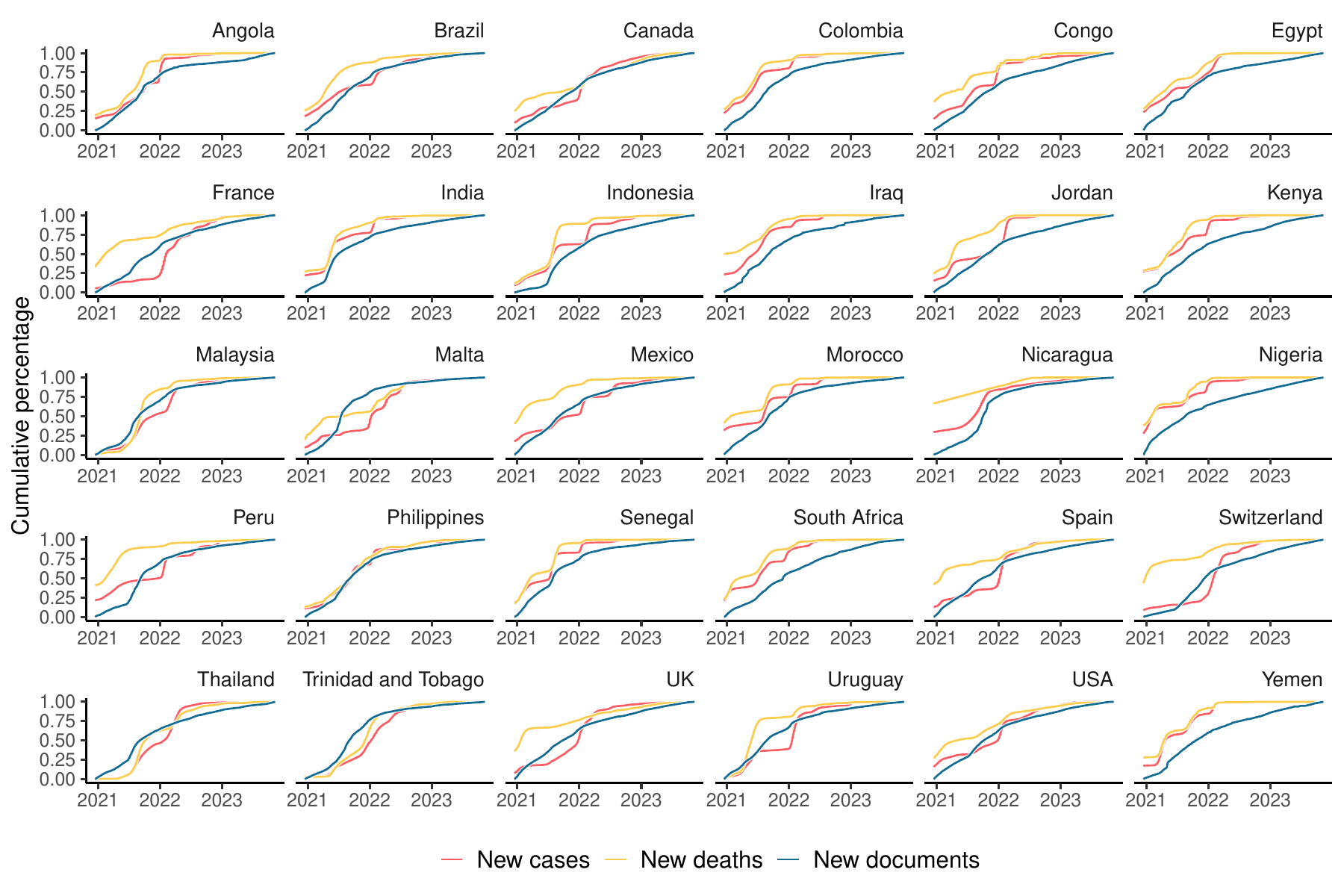}
\caption{{\bf Normalised cumulative curves of new cases, new deaths, and new documents.} 
Some countries exhibit coupling between the epidemiological curves and the production of new documents (e.g., Philippines, Malaysia), but in most cases the coupling is either partial (e.g., India) or absent (e.g., France).}
\label{fig:cumulative_vars}
\end{figure*}

In order to understand the potential association between the infodemic phenomenon, represented by the curve of new documents, and the epidemic phenomenon, represented by the curves of new cases and new deaths, we start our analysis with a comparison of these curves. 
Fig.~\ref{fig:cumulative_vars} shows the cumulative proportion of new cases, new deaths, and new documents over time. Overall, we note an association between the two epidemiological variables, which are only partially correlated to the cumulative production of documents. In some cases, we note coupling between the epidemiological and documents curves, as they tend to have peaks and plateaus (i.e., sudden rises or steady behaviours of the variables) at nearly the same time (see, for instance, India, Nicaragua, and the Philippines). In some other instances, new cases and new deaths are only partially coupled with new documents, showing a single common jump, as in the case of Indonesia. Otherwise, as in the cases of the UK, South Africa, and France, the curves tend to show a stronger decoupling, as the production of new documents is steadier than the evolution of the two epidemiological variables. For what concerns the case of France, we note, at the beginning of 2022, a marked peak in cumulative cases while the document curve maintains a steady trend; this can be interpreted as a signal of saturation in document production or even as a potential signal of information fatigue, a component of epidemic fatigue~\cite{lilleholt2023development}, according to which people and other actors refrained from posting about COVID-19, regardless of epidemic events. 
Another interesting instance occurs when a peak in the document curve is not associated with peaks in the epidemiological curves, as in the case of Malta. This situation may either indicate the enforcement of NPIs, which might fuel document production despite a steadiness of the epidemic curve, or a shift of focus in the national debate about the pandemic towards other countries, as observed on a regional to national scale in \cite{Castriota2023}. 

\subsection*{Modelling infodemic at a global scale}
\label{sec:infodemic_global}
Given the non-trivial nature of the coupling between document production and epidemiological variables displayed in Fig.~\ref{fig:cumulative_vars}, for which we provide a detailed correlation analysis in Supplementary Material (pp 2-6, Figures S1--S4), we employ a fixed effects regression model to understand and quantify the drivers behind the production of documents. 
In Table \ref{tab:regression_model1}, we report the coefficients obtained for all three models and their standard errors, which are robust to heteroskedasticity.

\begin{table}[!ht] 
    \centering 
  \caption{\textbf{Fixed-effect panel regression with robust standard errors}. The model includes dummy variables for the day of the week and season of the year.} 
  \label{tab:regression_model1} 
\begin{tabular}{@{\extracolsep{5pt}}lccc} 
\\[-1.8ex]\toprule 
\\[-1.8ex] & \multicolumn{3}{c}{New documents} \\ 
\\[-1.8ex] & (1a) & (1b) & (1c)\\ 
\midrule \\[-1.8ex] 
 New cases & 0.110$^{***}$ &  & 0.008\\ 
  & (0.019) &  & (0.021) \\ 
  & & & \\ 
 New cases (neighbours) & 0.159$^{***}$ &  & 0.007 \\ 
  & (0.016) &  & (0.022) \\ 
  & & & \\ 
 New deaths &  & 0.159$^{***}$ & 0.151$^{***}$ \\ 
  &  & (0.022) & (0.034) \\ 
  & & & \\ 
 New deaths (neighbours) &  & 0.258$^{***}$ & 0.248$^{***}$ \\ 
  &  & (0.022) & (0.030) \\ 
  & & & \\ 
\midrule \\[-1.8ex] 
Observations & 31,620 & 31,620 & 31,620 \\ 
R$^{2}$ & 0.400 & 0.493 & 0.493 \\ 
Adjusted R$^{2}$ & 0.399 & 0.493 & 0.493 \\ 
\bottomrule \\[-1.8ex] 
$^{*}$p$<$0.1; $^{**}$p$<$0.05; $^{***}$p$<$0.01\\ 
\end{tabular} 
\end{table} 

From models \eqref{eq:model_cases} and \eqref{eq:model_deaths}, we obtain positive and significant coefficients associated with the independent variables, which indicate an effect of epidemiological variables on new documents. When both variables are employed (model \eqref{eq:model_cases_and_deaths}), only new deaths are significant, likely due to the positive correlation between new cases and new deaths. Overall, the coefficients associated with such variables can be interpreted as the elasticity of the independent variable with respect to the dependent variable, thus providing a quantitative description of the extent to which a unit change in the evolution of new cases or new deaths affects the infodemic process in percentage points. 
For instance, the elasticity of new deaths being $\simeq0.16$ indicates that, within the model, a 1\% increase in new deaths is associated with a 0.16\% increase in document production. 
Furthermore, the property of regression coefficients being elasticity indexes allows for a comparison between models, as the values are independent of both the unit of measurement and the scale of the dependent and independent variables. With this in mind, the most interesting result concerns the relationship between the coefficients related to epidemiological variables in countries belonging to the same WHO region ($\beta_2$ and $\beta_4$) versus the domestic ones ($\beta_1$ and $\beta_3$): when the coefficients are significant, the foreign epidemic burden (measured either with new cases or new deaths) has a greater effect on document production than the domestic one. This can be interpreted as quantitative evidence for the fact that the COVID-19 infodemic was a phenomenon driven more by foreign epidemic burden than by the domestic one.

As a robustness check, we validate the overall effect of the epidemic variables on the information ecosystem by implementing a different model in which we substitute the dependent variable (i.e., new documents gathered by the WHO-EARS platform) with Google Trends data regarding COVID-19 (see Supplementary Material, Table S1). This model specification is particularly relevant in that we move from an information production (WHO-EARS) to an information demand (Google Trends) perspective, and obtain consistent results.

Finally, to quantify the extent to which other epidemiological variables affect the infodemic phenomenon, we define a regression model that also includes data regarding vaccine coverage and non-pharmaceutical interventions as regressors. 
Interestingly, the model achieves an adjusted $R^2$ of 0.43, which is slightly lower than that of the simpler model \eqref{eq:model_deaths}. However, consistently with our previous results, all included variables have a significant positive effect on the production of new documents, especially the percentage change in the vaccinated population (see Supplementary Material, Table S2).

\subsection*{Temporal sensitivity of the model}
\label{sec:temporal_sensitivity}
As it is reasonable to assume that the epidemic fuels the infodemic consistently throughout its development, we now aim to investigate whether our model reflects the possibly weaker relationship between the phenomena towards the end of the health crisis. From a technical point of view, such variations in the association between the dependent and independent variables over time would be highlighted by changes in the elasticity indexes. Indeed, while the coefficient values for the general model are fixed, we expect them to change when fitting the same model over different time windows over the time frame under study. To this end, we evaluate the temporal sensitivity of the model by studying how the elasticity changes over time using model~\eqref{eq:model_deaths}, as it is the specification that better explains the data in terms of $R^2$.

\begin{figure*}[!t]
\centering
\includegraphics[width =1\linewidth]{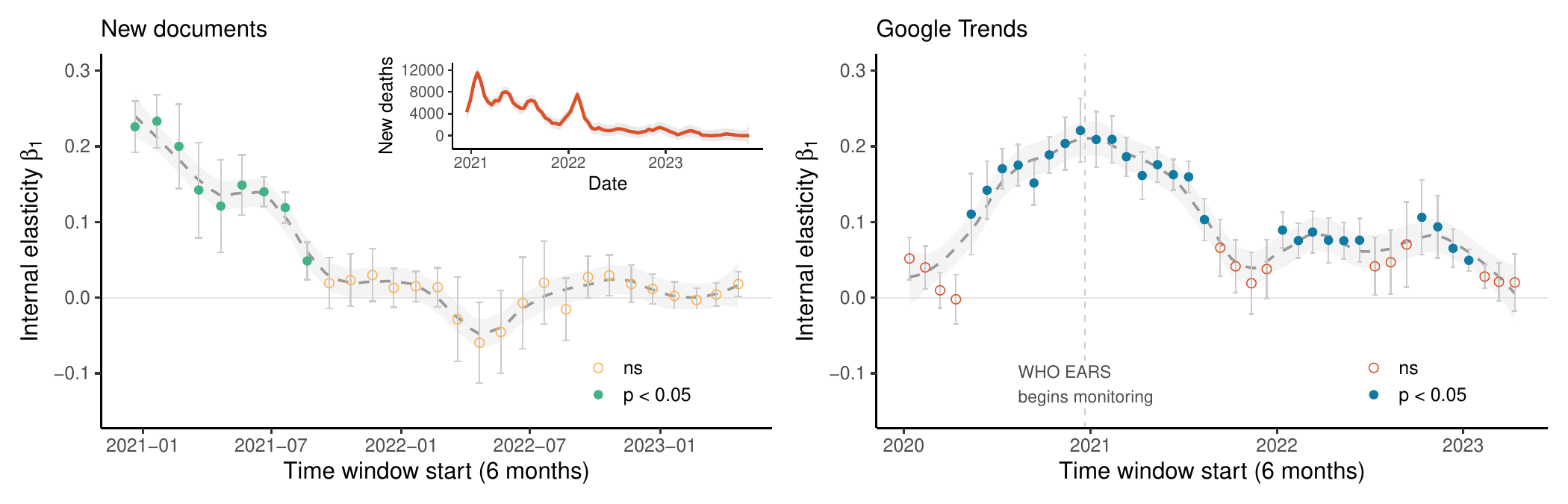}
\caption{
    \textbf{Temporal evolution of the internal elasticity associated with the number of new deaths.} We fit the panel regression model specified in \eqref{eq:model_deaths} over 6-month sliding windows, and display how the internal elasticity $\beta_1$ and its robust standard error evolve. Left, the evolution of $\beta_1$ using the number of new documents as the dependent variable, with the inset reporting the total number of new deaths across all modelled countries. Right, the evolution of $\beta_1$ using Google Trends data as the dependent variable, as a proxy for information demand rather than production. Both curves show similar behaviour and comparable values, peaking in the first half of 2021 and stabilizing around values close to 0 from the end of 2021 onward.
}
\label{fig:internal_elasticity}
\end{figure*}

To substantiate this reasoning, in Fig.~\ref{fig:internal_elasticity} (left panel), we report the evolution of the internal elasticity $\beta_1$, which encodes the responsiveness of the infodemic to domestic changes in the new death counts. In detail, we create rolling windows of 6 months and employ model specification~\eqref{eq:model_deaths}.
The resulting curve is characterised by a decreasing trend: we observe a peak during the first half of 2021, followed by oscillations around zero starting from the end of 2021. A potential explanation behind this result could be the rollout of vaccines and the associated volume of digital content about them, which might have strengthened the impact of the epidemic burden on the infodemic. However, due to the complexity of the infodemic phenomenon and the aggregated nature of our data, it is difficult to attribute this to a single factor, and alternative interpretations remain plausible.

Finally, as the WHO-EARS data collection only began at the end of 2020, we perform a robustness check including the year 2020 but changing the dependent variable, as reported in Fig. \ref{fig:internal_elasticity} (right panel). The new dependent variable shifts the focus towards information demand and is represented by a collection of COVID-19 data from Google Trends. Our findings highlight not only the same decreasing trend of the elasticity coefficient from the first half of 2021 onward, but also an increasing trend during 2020 that reaches the same peak in the first 6 months of 2021 (see Supplementary Material, Figure S5 for the overlapped curves). We perform an additional robustness check that involves using a rolling window of 3 months rather than 6 months. The resulting internal elasticity curves, reported in Supplementary Material Figure S6, show a broadly consistent evolution with those shown in Fig. \ref{fig:internal_elasticity} for both new documents and Google Trends data.

\subsection*{Country-level differences in infodemic modelling}
\label{sec:country_level}
After modelling and examining the epidemic–infodemic relationship globally, a natural step is to study the infodemic response of single countries. This kind of analysis provides additional insights into how national infodemics may deviate from the general trend. 

\begin{figure*}[!t]
\centering
\includegraphics[width = 0.7\linewidth]{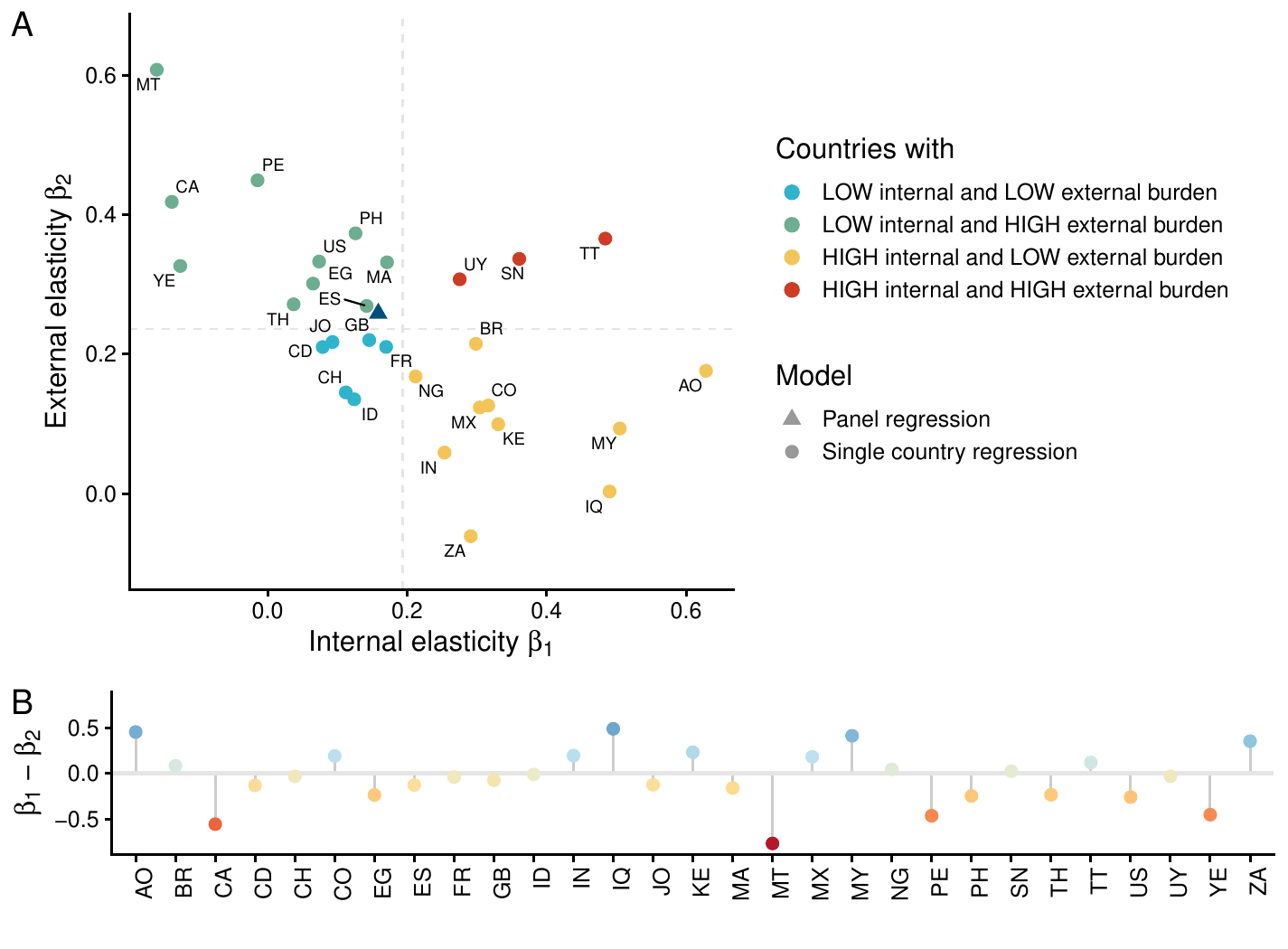}
\caption{
    \textbf{Country-specific internal and external elasticities associated with the number of new deaths.} For each country, we fit a regression model as specified in \eqref{eq:model_deaths} and obtain an estimate of the internal elasticity $\beta_1$ and of the external elasticity $\beta_2$. (A) Comparison of $\beta_1$ and $\beta_2$ for all countries, with the vertical and horizontal dashed lines reporting the average $\beta_1$ and $\beta_2$, respectively, and the triangle indicating the values resulting from the panel regression. (B) Difference between internal and external elasticity $\beta_1 - \beta_2$ for all countries. Nicaragua was omitted from this figure because its estimated $\beta_1$ was much higher than that of all other countries. 
}
\label{fig:internal_v_external}
\end{figure*}

We perform a regression analysis for each country, and subsequently study the internal $\beta_1$ and external $\beta_2$ elasticity, where the former represents the elasticity index obtained from domestic epidemiological variables and the latter represents the elasticity index obtained from epidemiological variables of countries in the same WHO region. Fig.~\ref{fig:internal_v_external} panel A displays the relationship between internal and external elasticity indexes across countries, showing heterogeneous results. Using the mean values of external and internal elasticity indexes as references, we divide the graph into four quadrants that provide a data-driven classification of country-level response, based on the relationship between internal and external elasticity. 
While most countries cluster near the junction between the axes marked by the average values, we note some exceptions. For instance, Canada (CA), located in the upper left quadrant, exhibits a relatively large external elasticity but an internal elasticity close to 0, suggesting that the evolution of new deaths in neighbouring countries (likely, the US) had a larger impact on document production than the domestic ones. Conversely, Iraq (IQ), located in the lower right quadrant, is characterised by a large internal elasticity and an external elasticity of approximately 0, showing a lower impact of epidemic data from neighbouring countries on the national infodemic. Fig.~\ref{fig:internal_v_external} panel B displays the difference between internal and external elasticity indexes for each country. In countries that show a difference above 0, the domestic epidemic burden has a higher impact on the infodemic compared to the burden from neighbouring countries. In countries with a difference value below 0, the infodemic is mostly fuelled by the foreign epidemiological situation (proportionally to the magnitude of the difference), while the impact of the domestic epidemic burden on document production is limited.

All estimated values of $\beta_1$ and $\beta_2$ are fully reported in Supplementary Material Figure S7, including the two resulting from the panel regression with all countries.

\section*{Discussion}
\label{sec:discussion}
In this paper, we quantitatively analyse how epidemic variables drive the infodemic, using data-driven models based on diverse official sources, including the WHO, the University of Oxford, and Google. The study incorporates data on COVID-19 cases, related deaths, vaccine coverage, and non-pharmaceutical interventions, as well as information production and demand.
This approach leads to key insights regarding the infodemic phenomenon that can inform public health measures, infodemic management strategies, and, potentially, infodemic prediction models. 
Our models show a clear relationship between document production and epidemiological variables (i.e., the COVID-19 infodemic and the epidemic), with the number of new cases and new deaths being significant drivers of the infodemic. Moreover, we show that new deaths outperform other epidemic-related variables as infodemic predictors. The relevance of this variable resonates with recent findings from social psychology, according to which the salience of death brought on by COVID-19 played a central role in driving the attitudes and behaviour of the population~\cite{pyszczynski2021terror}. Regarding the epidemic variables, one of the most significant findings is that the epidemiological situation in neighbouring countries emerges as a strong predictor of the infodemic at the national level. If we interpret the foreign epidemic situation as a measure of pandemic proximity, we can speculate that the fear of impending domestic contagions was a significant driver of the infodemic phenomenon. This highlights the critical importance of coordinated, cross-border efforts to mitigate the spread of misinformation and disinformation during health crises.

From a technical perspective, the coefficients in our model represent elasticity factors, providing a robust and scale-independent measure of how percentage changes in independent variables affect the dependent variable (i.e., the intensity of the infodemic). Furthermore, the model displays temporal sensitivity with respect to the epidemic burden and thus has more explanatory power when fitted on the early phases of the epidemic/pandemic. 
This aspect of the model highlights its versatility and accuracy in providing robust results that can be used in scenarios where the independent variable is subject to sudden changes. Furthermore, a model characterised by such fine temporal sensitivity enables us to develop reasonable scenarios of infodemic trends by accounting for variations in the effects of different independent variables across different time windows, aligning with common practices in epidemic management~\cite{jewell2020predictive}. The results are robust to a stress test performed using Google Trends data as an alternative dependent variable~\cite{Nghiem2016googletrends, Gravino2022supply}, therefore considering information demand as a proxy for the infodemic. This dual perspective aligns closely with a broader definition of infodemics, which encompasses both the generation and consumption of information~\cite{health2024rethinking}.

\section*{Limitations of the Study}
\label{sec:limitations}
Our study has a number of limitations. First, data on information production are available only after December 2020, and this might result in a partial image of the whole information dynamic. Secondly, despite considering 30 countries, for certain areas of the world, such as Eastern Europe, Russia, China, and Australia, data are not available, and therefore a geographical bias could have been introduced in the model, limiting its accuracy and generalisability. Relatedly, our definition of ``neighbouring countries'' based on WHO regions may only partly reflect actual information networks, since cross-border effects can arise from the interaction of several complex mechanisms, such as linguistic, cultural, and media ties, that are not captured by geographical proximity alone. At the same time, many of these mechanisms are inherently difficult to quantify and incorporate into a model without sacrificing its interpretability. Lastly, we considered only document production and information needs as proxies for the infodemic, therefore excluding some other crucial measures of the infodemic, such as information voids \cite{Scalco2026} and disinformation \cite{Wang2026}.
This methodological choice reflects the absence of a consensus on the operational definition of an infodemic. Consequently, we relied on the WHO’s core concept of ``information overabundance'' and adopted document production as our proxy.

In addition to these data constraints, the continuous modelling approach adopted in this study does not provide a discrete label or classification for the observed infodemic. Indeed, a major shortcoming in the current literature is the lack of an operational definition for overabundance, the main concept that underlies the WHO’s definition of an infodemic. Robustly classifying infodemic regimes would require establishing a formal baseline model capable of quantifying a metrics-based excess information, conceptually analogous to the excess mortality metric utilized in epidemiology. However, defining such a baseline remains far from trivial due to persistent challenges in data collection, structural information uncertainty, and the dynamic mapping of online ecosystems. Developing standardized baselines to empirically bound infodemic regimes thus represents a critical avenue for future research.
Finally, a structural limitation of this study lies in the aggregation of multi-platform data as provided by the WHO-EARS platform. Because an infodemic is intrinsically a multi-platform phenomenon, analyzing data aggregated across channels may confound platform-specific dynamics, user behaviors, or echo platform effects that a disaggregated analysis could otherwise reveal \cite{Iqhrammullah2025}. However, our choice of data was dictated by the need for source consistency, that is, matching infodemic metrics directly with the official country-level epidemiological data provided by the World Health Organization (WHO). Future research should aim to integrate multi-platform, disaggregated datasets to evaluate how individual digital ecosystems contribute to the broader, macro-level infodemic surges observed here.

\section*{Conclusions}
\label{sec:conclusions}
The findings from this study hold significant implications for the management of future infodemics. The influence of neighbouring countries' epidemiological situations underscores the necessity of international cooperation to address the inherently global nature of infodemics. Information spread transcends borders, requiring transnational strategies to counteract misinformation effectively.
Our results demonstrate that the evolution of infodemics and their relationship with epidemic variables can be closely monitored, even over short time windows.
Just as epidemic models allow for the evaluation of different epidemic management strategies, our model enables the estimation and potential prediction of the effects of various infodemic management measures on the infodemic burden. The model's robustness in capturing both information demand and production creates an opportunity for developing early-warning systems to detect and mitigate emerging infodemics promptly.
Future specifications of the model could better capture the complexity of the infodemic, enhancing both its accuracy and explanatory power. Assuming some degree of similarity between an infodemic and a syndemic~\cite{singer2017syndemics, horton2020offline}, where a multitude of interacting factors, such as biological, social, and geopolitical ones, play a significant role, additional variables should be considered. In the considered case, variables such as education level, economic indicators, and internet access are more readily available and could provide valuable insights if included in future analyses. Others, such as health and digital literacy, trust in media and institutions, and characteristics of the media landscape, are potentially harder to measure but could be explored through regional investigations, especially with greater resource allocation. Additionally, combining measures of document production and information demand into a single variable could offer a more integrated perspective on the interplay between these two dimensions of the infodemic. Furthermore, variables like the estimation of information voids and the prevalence of circulating misinformation and disinformation, while requiring advanced computational capabilities, hold great potential for broadening the scope of infodemic modelling. 
Including these factors would provide a more comprehensive framework for understanding and addressing the multifaceted dynamics of infodemics and enable policymakers and health organizations to identify and respond to emerging infodemics more effectively, preventing escalation.
Overall, this study provides a comprehensive framework for understanding the dynamics of infodemics and offers actionable insights for developing more effective containment strategies in the future. While the model already provides valuable information, it also has the potential for further refinement by incorporating additional variables and perspectives to enhance its explanatory power and widen its applicability. 

\section*{Materials and methods}
\label{sec:mats_and_methods}

\subsection*{Data}
\label{sec:data}

In this section, we describe the publicly available datasets that we employed in our analyses. A complete breakdown of the collected data is reported in Table~\ref{tab:data_summary}.
\begin{table*}[ht]
\small
    \centering
    \begin{tabular}{llccc}
        \toprule
        Data & Source & N. Countries & Time Window (yy-mm-dd) & Time Scale\\
        \midrule
        New Cases & WHO & 236 & 2020-01-03 -- 2023-11-09 & Daily\\
        New Deaths & WHO & 236 & 2020-01-03 -- 2023-11-09 & Daily\\
        New Documents & WHO-EARS & 30 & 2020-12-15 -- 2023-11-09 & Daily\\
        Vaccinations (\%) & OxCGRT & 183 & 2020-01-01 -- 2023-02-28 & Daily\\
        Stringency Index & OxCGRT & 183 & 2020-01-01 -- 2023-02-28 & Daily\\
        Google Trends & Google & 30 & 2020-01-05 -- 2023-11-05 & Weekly \\
        \bottomrule
    \end{tabular}
    \caption{Summary of variables used for data analysis.}
    \label{tab:data_summary}
\end{table*}
We employed the WHO Coronavirus (COVID-19) surveillance data that reports new COVID-19 cases and deaths on a daily scale. The dataset provides information for 236 countries starting from 2020-01-03. The data, displayed in a dashboard~\cite{whoCOVID19Cases}, are freely available~\cite{whoCOVID19Data}. 
The version of the dataset used for the present study tracks the number of new cases and new deaths until 2023-11-09 and can be retrieved from the Wayback Machine~\cite{dataWaybackMachine}.

Although the infodemic encompasses a broad conceptualization, which includes behavioural, social, and health system dynamics, as highlighted by the WHO~\cite{wilhelm2023measuring}, in this study we focused on information production to develop explanatory models. Documents related to COVID-19 were collected on a daily basis by the WHO Early AI-supported Response with Social Listening Platform (EARS). Documents include public conversations and public social media content related to COVID-19, collected using the Twitter API, and data aggregators of public sources such as forums, message boards, blogs, and comments in news. The dataset provides information for 30 countries starting from 2020-12-15 and can be found in the official WHO EARS repository~\cite{dataWhoEars}. Additionally, we collected Google Trends data related to COVID-19 for the same period and set of countries encompassed by the previously described epidemiological data, to serve as a representative metric of information demand, rather than production.

We used the Oxford Covid-19 Government Response Tracker (OxCGRT)~\cite{hale2021global}, a publicly available dataset~\cite{dataOxCGRTcovidpolicydataset} collecting information on the measures enacted by a government in response to the pandemic, to retrieve the status of COVID-19 vaccination for each country. More in detail, the dataset keeps track of the daily proportion of the vaccinated population in 183 countries from 2020-01-01 to 2023-02-28.

\subsection*{Covariates}
\label{sec:covariates}
From the WHO COVID-19 surveillance data, we derived four covariates related to the epidemiological situation. Two of these represent a country's domestic epidemiological burden: the 7-day right-aligned rolling average of the number of new daily cases and new daily deaths, for each country. The other two covariates capture the regional epidemiological context: the 7-day right-aligned rolling average of the number of new daily cases and new daily deaths in neighbouring countries. Note that we defined two countries as neighbours if they belong to the same WHO region.

Further, from the OxCGRT data, we derived two additional covariates that may also affect the infodemic phenomenon. One corresponds to the daily increase in the percentage of the vaccinated population in a country. The other is the Stringency Index (SI), a composite measure that quantifies the extent of containment and closure measures adopted by a government based on nine indicators, including school and workspace closings, cancellation of public events, restrictions on gatherings, travel controls, and stay-at-home orders. The SI ranges between 0 and 100, where 100 indicates the maximum stringency of non-pharmaceutical interventions, such as a full lockdown. We also considered the 7-day right-aligned rolling average for these two covariates.

\subsection*{Statistical Analysis}
\label{sec:stat_analysis}

We employed a fixed effects regression model to understand and quantify the drivers behind the production of documents (infodemics data). To this end, we utilised a panel regression framework with fixed effects (i.e., separate intercept values that encode country-specific effects) where the daily number of new documents is treated as the dependent variable $y$. Specifically, we considered the 7-day right-aligned rolling average of new daily documents, consistent with the approach used for the covariates. Additionally, we log-transformed both independent and dependent variables. The panel data are balanced (i.e., each panel contains the same number of observations) and include a total of $N=30$ countries with $T=1054$ observations each (from 2020-12-21 to 2023-11-09).
We specified model \eqref{eq:model_cases} as follows:
\begin{equation}
\label{eq:model_cases}
\tag{1a}
y_{it} = \alpha_i + \sum_{j=1}^6 \gamma_j d_{j,t}^{\text{day}} + \sum_{j=1}^3 \delta_j d_{j,t}^{\text{season}} + \beta_1 C_{it} + \beta_2 \overline{C}_{it} + \epsilon_{it} 
\end{equation}
where $\alpha_i$ is the time-invariant fixed effect for country $i$, $d_{j,t}^{\text{day}}$ are dummy variables for the weekdays, and $d_{j,t}^{\text{season}}$ for the seasons of the year. The number of new cases in country $i$ at time $t$ is denoted with the variable $C_{it}$, whereas we denote with $\overline{C}_{it}$ the total number of new cases at time $t$ in all other countries in the same WHO region of country $i$. Similarly to model \eqref{eq:model_cases}, we also defined a second specification \eqref{eq:model_deaths} where we used the number of deaths $D_{it}$ as the epidemiological independent variable
\begin{equation}
\label{eq:model_deaths}
\tag{1b}
y_{it} = \alpha_i + \sum_{j=1}^6 \gamma_j d_{j,t}^{\text{day}} + \sum_{j=1}^3 \delta_j d_{j,t}^{\text{season}} + \beta_1 D_{it} + \beta_2 \overline{D}_{it} + \epsilon_{it} 
\end{equation}
and a third specification \eqref{eq:model_cases_and_deaths} where both cases and deaths were used as independent variables
\begin{equation}
\label{eq:model_cases_and_deaths}
\tag{1c}
\begin{aligned}
y_{it} = &\alpha_i + \sum_{j=1}^6 \gamma_j d_{j,t}^{\text{day}} + \sum_{j=1}^3 \delta_j d_{j,t}^{\text{season}}\\
&+ \beta_1 C_{it} + \beta_2 \overline{C}_{it} + \beta_3 D_{it} + \beta_4 \overline{D}_{it} + \epsilon_{it}
\end{aligned}
\end{equation}

We also defined a regression model that includes data regarding vaccine coverage and non-pharmaceutical interventions as regressors. In particular, we added to model specification \eqref{eq:model_deaths} a new covariate $V_{it}$ that represents the change from time $t-1$ to time $t$ of the percentage of the vaccinated population in country $i$, and a covariate $S_{it}$ that is the Stringency Index at time $t$ for country $i$. This regression model was thus specified as follows:
\begin{equation}
\label{eq:model_all}
\tag{1d}
\begin{aligned}
y_{it} = &\alpha_i + \sum_{j=1}^6 \gamma_j d_{j,t}^{\text{day}} + \sum_{j=1}^3 \delta_j d_{j,t}^{\text{season}}\\
&+ \beta_1 D_{it} + \beta_2 \overline{D}_{it} + \beta_3 V_{it} + \beta_4 S_{it} + \epsilon_{it} 
\end{aligned}
\end{equation}
To make sure that the potential multicollinearity of the epidemiological variables with the newly added regressors had no impact on the results, we computed the Variance Inflation Factor (VIF) for model \eqref{eq:model_all}, obtaining a $\text{VIF} \lessapprox 2$ (i.e., non-concerning multicollinearity) for all independent variables.

\section*{Funding}
F.G. acknowledges project CRESP from the Italian Ministry of Health under the program CCM 2022.

\section*{Declaration of interests}
F.G. received speaker honorarium from MSD Italia and Moderna.

\section*{Author contributions statement}
\textbf{Edoardo Loru}: Conceptualization, Formal analysis, Investigation, Methodology, Software, Writing - original draft, Writing – review \& editing, Visualization. 
\textbf{Francesco Gesualdo:} Conceptualization, Writing – original draft, Writing – review \& editing.
\textbf{Marco Delmastro}: Conceptualization, Methodology. 
\textbf{Matteo Cinelli}: Conceptualization, Formal analysis, Investigation, Methodology, Software, Writing – original draft, Writing – review \& editing, Supervision.

\section*{Data availability}
All data underlying this article are publicly available. Data for new COVID-19 cases and deaths can be retrieved from the WHO \cite{dataWaybackMachine}. Social listening data from WHO-EARS is available on a GitHub repository \cite{dataWhoEars}. Data on vaccinations and on the Stringency Index is available on the OxCGRT GitHub repository \cite{dataOxCGRTcovidpolicydataset}. Google Trends data can be freely collected from the official web interface (\href{https://trends.google.com}{https://trends.google.com}).

\bibliography{sn-bibliography}% common bib file

\backmatter

\end{document}